\documentclass[%
superscriptaddress,
twocolumn,
 amsmath,amssymb,
 aps,
prb,
]{revtex4-1}

\usepackage{graphicx}
\usepackage{dcolumn}
\usepackage{bm}
\usepackage{hyperref}

\renewcommand{\v}[1]{\ensuremath{\mathbf{#1}}} 
\newcommand{\gv}[1]{\ensuremath{\mbox{\boldmath$ #1 $}}}
\newcommand{\pd}[2]{\frac{\partial #1}{\partial #2}}

\newcommand{\grad}[1]{\gv{\nabla} #1} 
\renewcommand{\div}[1]{\gv{\nabla} \cdot #1} 
\newcommand{\curl}[1]{\gv{\nabla} \times #1} 
\let\baraccent=\= 
\renewcommand{\=}[1]{\stackrel{#1}{=}} 

\begin{document}

\title{Transmission/reflection coefficients and Faraday/Kerr rotations as a function of applied magnetic fields in spin-orbit coupled Dirac metals}

\author{Jinho Yang, Jeehoon Kim, and Ki-Seok Kim}

\affiliation{Department of Physics, Pohang University of Science and Technology, Pohang 790-784, Korea}

\date{\today}

\begin{abstract}
We reveal the nature of propagation and reflection of light in spin-orbit coupled Dirac metals under external magnetic fields. Such applied magnetic fields split the four-fold degeneracy of a spin-orbit coupled Dirac metal state into a pair of a two-fold degeneracy along the direction of the applied magnetic field, resulting in a Weyl band structure. These Weyl metals turn out to play the role of a chiral prism, whose electromagnetic properties are described by axion electrodynamics: An incident monochromatic wave can split into three differently polarized modes (eigenvectors), propagating with different wave numbers (eigenvalues). In particular, the axion electrodynamics allows the longitudinal component naturally inside the Weyl metal state. We evaluate both transmission/reflection coefficients and Faraday/Kerr rotation angles as a function of both an external magnetic field and frequency for various configurations of light propagation. The helicity of the propagating/reflected light is determined by $\bm{\nabla} \theta \times \bm{E}_{light} = \bm{B}_{ext} \times \bm{E}_{light}$, where $\bm{\nabla} \theta = \bm{B}_{ext}$ is the gradient of the $\theta-$field in the axion term given by the applied magnetic field and $\bm{E}_{light}$ is the electric-field of the incident light. This implies that the direction of the external magnetic field controls the Faraday/Kerr rotation. We find several interesting optical properties of the Weyl metal phase. First, longitudinal oscillating charge-density fluctuations along the light propagating direction arise when the pair of Weyl nodes are aligned along the direction of the oscillating magnetic field, which give rise to the longitudinal component of the electromagnetic wave. Second, the Weyl metal phase becomes more reflective when the external magnetic field is enhanced to be along with $\v{E}//\bm{B}_{ext}$ due to the longitudinal negative magnetoresistivity, which is a fingerprint of the Weyl metal phase. Third, eigenmodes can have various structures, depending on a parameter $\eta$, which corresponds to a ratio between the conventional Hall effect from normal electrons and the anomalous Hall effect from Weyl electrons. We propose these strong magnetic-field dependencies of the optical response as the fingerprints of the axion electrodynamics.
\end{abstract}

\maketitle

\section{Introduction}

Anomalous transport phenomena in Weyl metals \cite{WM1,WM2,WM3,WM_Reviews}, resulting from effects of both Berry curvature and chiral anomaly \cite{Nielsen_Ninomiya}, have been investigated extensively in both experimental and theoretical aspects. In particular, both anomalous Hall and chiral magnetic effects have been discussed theoretically based on the Boltzmann transport theoretical framework modified by the Berry curvature and the chiral anomaly \cite{CME1,CME2,CME3,CME4,CME5,CME6,CME7, Boltzmann_Chiral_Anomaly1,Boltzmann_Chiral_Anomaly2,Boltzmann_Chiral_Anomaly3,Boltzmann_Chiral_Anomaly4,Boltzmann_Chiral_Anomaly5,Boltzmann_Chiral_Anomaly6,Boltzmann_Chiral_Anomaly7, Boltzmann_Chiral_Anomaly8,AHE1,AHE2,AHE3}. The so called negative longitudinal magnetoresistivity has been observed experimentally in various types of Weyl metals \cite{NLMR_First_Exp,NLMR_Followup_I,NLMR_Followup_II,NLMR_Followup_III,NLMR_Followup_IV, NLMR_Followup_V,NLMR_Followup_VI,NLMR_Followup_VII,NLMR_Followup_VIII}, also well understood by the topologically ``generalized" Boltzmann transport theory \cite{CME1,CME2,CME3,CME4,CME5,CME6,CME7, Boltzmann_Chiral_Anomaly1,Boltzmann_Chiral_Anomaly2,Boltzmann_Chiral_Anomaly3,Boltzmann_Chiral_Anomaly4,Boltzmann_Chiral_Anomaly5,Boltzmann_Chiral_Anomaly6,Boltzmann_Chiral_Anomaly7,Boltzmann_Chiral_Anomaly8}. However, not only the Boltzmann transport theory but also the Maxwell's electrodynamics theory should be modified by such topological ingredients. The chiral-anomaly modified Maxwell theory is referred to as the theory of axion electrodynamics, where an emergent $\bm{E} \cdot \bm{B}$ term is introduced into the conventional Maxwell action of electromagnetic fields \cite{Axion_EM}. The angle coefficient of the $\bm{E} \cdot \bm{B}$ term turns out to be a function of space and time, originating from an effective Weyl band structure \cite{Boltzmann_Chiral_Anomaly8,AHE1,AHE2,Disordered_Weyl_Metal1,Disordered_Weyl_Metal2,Anomaly_ISBWM}. Unfortunately, both theoretical and experimental studies for this axion electrodynamics have not been performed satisfactorily yet. In particular, light scattering measurements are still missing as far as we know, where the axion electrodynamics theory in Weyl metals with broken time reversal symmetry has not been proven yet.

In this study we investigate both transmission/reflection coefficients and Faraday/Kerr rotation angles in spin-orbit coupled Dirac metals, particularly, as a function of both the external magnetic field and the frequency of light for various configurations of light propagation. It is well known that the applied magnetic field splits the four-fold degeneracy into a pair of the two-fold degeneracy, which gives rise to a Weyl band structure. Based on the chiral anomaly calculation, the $\theta$ coefficient of the topological-in-origin $\bm{E} \cdot \bm{B}$ term has been found, the gradient of which is given by the distance of the pair of Weyl points, resulting from the applied magnetic field \cite{AHE1,AHE2,Anomaly_ISBWM}. As a result, electromagnetic properties of spin-orbit coupled Dirac metals under external magnetic fields are governed by the axion electrodynamics. Our theoretical predictions on the magnetic-field dependence, more precisely, the gradient $\theta$ dependence, can be regarded as a guideline for experiments, proposed to be one of the fingerprints of the axion electrodynamics in Weyl metals.

Before going into our work, we would like to discuss recent studies related with ours. Faraday/Kerr rotation angles have been measured as a function of both frequencies and external magnetic fields for the surface state of a topological insulator, where anomalous Hall effects associated with the axion electrodynamics give rise to such rotations \cite{Axion_EM_Exp_TI_I,Axion_EM_Exp_TI_II}. Surface plasmon modes were also observed at the interface between topologically non-trivial cylindrical core and topological-trivial surrounding material, involved with the axion electrodynamics and modified constitutive relations \cite{Axion_EM_Exp_TI_III}. Various interesting theoretical studies of electromagnetic waves in a Weyl metallic state have been performed during recent few years \cite{AxionEMreferee1,AxionEMreferee2,AxionEMreferee3,Axion_EM_Th_WM,AxionEMreferee4}. Especially, transmission/reflection coefficients and Faraday/Kerr rotations in the axion electrodynamics have been discussed in Refs. \cite{AxionEMreferee1,AxionEMreferee2,AxionEMreferee3,Axion_EM_Th_WM}. We would like to point out that our results differ from these previous studies in several aspects although some parts are all consistent with each other; i) our normal modes inside the Weyl metal phase differ from those of the previous studies when the pair of Weyl nodes are aligned along the direction of the oscillating magnetic field, ii) reflectivity enhancement in the case of $\v{E}//\v{B_{ext}}$ results from the longitudinal negative magnetoresistance, iii) we find eigenmodes depending on the parameter $\eta$, which is the ratio between the conventional Hall conductivity from normal electrons and the anomalous Hall conductivity from Weyl electrons, and iv) external magnetic-field dependencies on both transmission/reflection coefficients and Faraday/Kerr rotation angles are the fingerprints of the Weyl metal phase.

\section{Axion electrodynamics in Weyl metals with broken time-reversal symmetry \label{axion electrodynamics in weyl metals} \label{generalsol}}

\subsection{Axion electrodynamics \label{Axionem}}

\begin{figure}
\centering
\includegraphics[width=8cm]{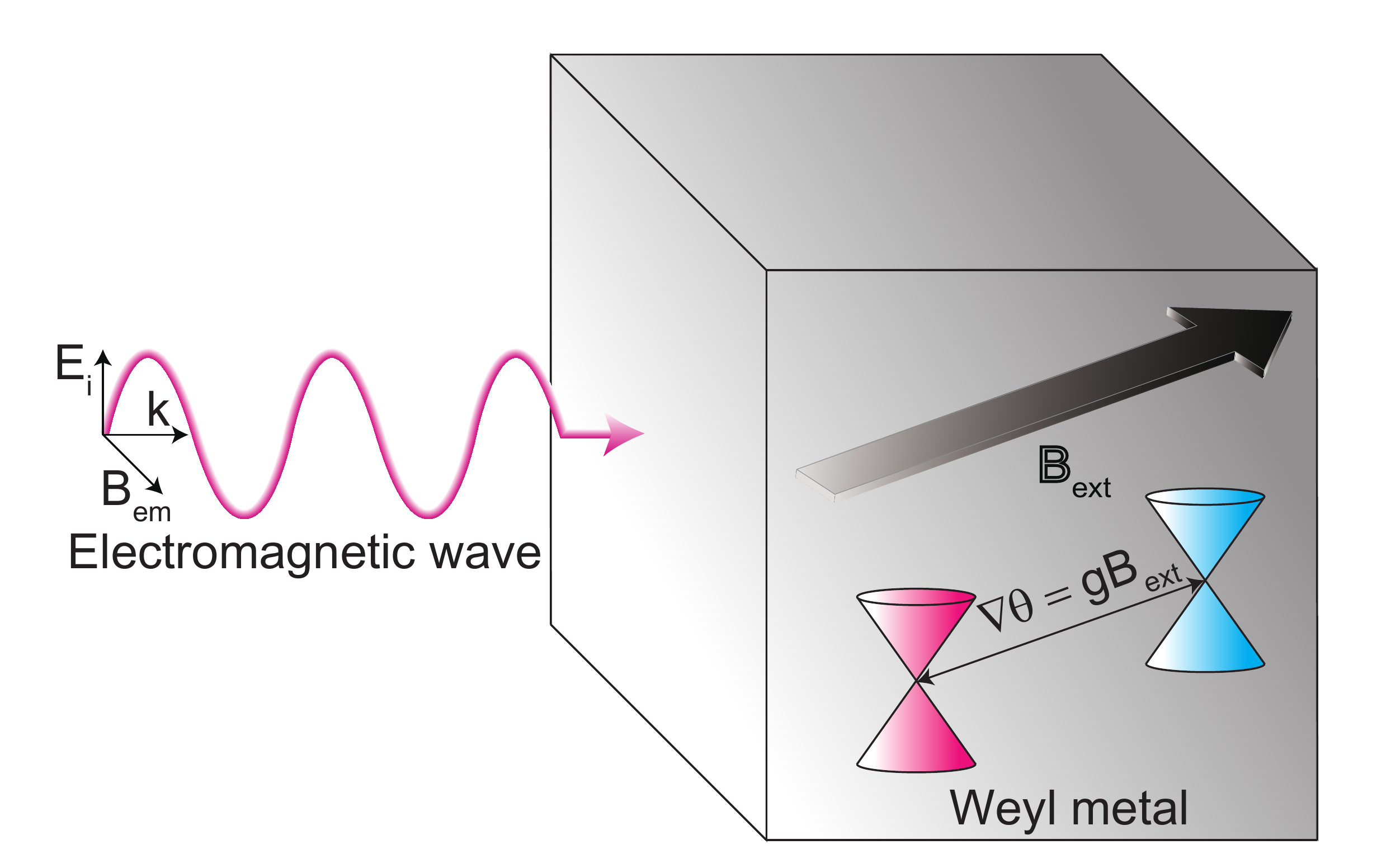}
\caption{Schematic diagram of the physical setup}
\label{incidenttoweyl}
\end{figure}

Our theoretical setup is shown in Fig. \ref{incidenttoweyl}. The setup has similarity to an optical response of a ferromagnetic system. Indeed, some optical properties such as the Faraday/Kerr rotation have resemblance to a ferromagnetic system (see Fig. \ref{eliptic}a). However, there exist significant differences between two systems because their origins of anomalous electromagnetic responses are totally different. For instance, the anomalous Hall conductivity ($\sigma_{\text{Hall}}$) in a ferromagnetic system mainly originates from three types of scattering mechanisms of conduction electrons due to magnetic moments \cite{AHE_Review} and the Faraday/Kerr rotation is mainly due to non-diagonal magnetic permeability tensor. However, in Weyl metals, the anomalous Hall conductivity ($\sigma_{\text{Weyl}}$) and the Faraday/Kerr rotation both originate from $\grad \theta$, which is linearly proportional to the momentum space distance between a Weyl pair. Therefore, not only the magnetic field dependence of the Hall effect should be different from each other, but also the Faraday/Kerr rotation and electromagnetic (EM) mode oscillating in the longitudinal direction exists even with the diagonal magnetic permeability and electric permittivity in a Weyl metal state as described below.

Now, we are going to present how $\grad \theta$ makes the anomalous Hall conductivity ($\sigma_{\text{Weyl}}$) and the Faraday/Kerr rotation by solving the modified Maxwell equation in a Weyl metal phase. An incident beam propagating along the $\bm{\hat{z}}-$direction is shined homogeneously on an infinite $xy$ plane of a Weyl metal under an arbitrary external magnetic field, $\v{B_\text{ext}} = (B_x,B_y,B_z)$. Here, the Weyl-metal sample is semi-infinite in the $\bm{\hat{z}}-$direction. Electromagnetic properties of the Weyl metal state are described by axion electrodynamics \cite{Axion_EM}
\begin{eqnarray}
\div \textbf{D} &=& \rho + \frac{2 \alpha}{\pi} \sqrt{\frac{\epsilon_0}{\mu_0}} \grad \theta \cdot \textbf{B} \label{Maxwell1} \\
\div \textbf{B} &=& 0 \label{Maxwell2}\\
\curl \v{E} &=& -\pd{\v{B}}{t} \label{Maxwell3}\\
\curl \v{H} &=& \pd{\v{D}}{t} + \v{J} -\frac{2\alpha}{\pi}\sqrt{\frac{\epsilon_0}{\mu_0}} \grad{\theta} \times \v{E} ,
\label{Maxwell4}
\end{eqnarray}
modified by the appearance of a position-dependent $\theta-$term from the Maxwell dynamics. Actually, one finds $\grad \theta = g \v{B}$ with a Lande$-g$ factor, where $2 \grad \theta$ gives the momentum space distance between a pair of Weyl points \cite{AHE1,Disordered_Weyl_Metal1,Disordered_Weyl_Metal2}. Here, we resort to $\v{D} = \epsilon \v{E}$ and  $\v{B} = \mu \v{H}$, where homogeneous permittivity $\epsilon$ and permeability $\mu \approx \mu_0$ are taken into account. For simplicity, we replace $\frac{\alpha}{\pi}\sqrt{\frac{\epsilon_0}{\mu_0}}$ with $\alpha$.

Taking curl to Eq. (\ref{Maxwell3}), we obtain
\begin{eqnarray}
\curl (3) &=& \curl \curl \v{E} \nonumber \\
              &=& \grad (\div \v{E}) - \grad^2 \v{E} \nonumber \\
              &=& \grad (\frac{\rho}{\epsilon} + \frac{2 \alpha}{\epsilon}) - \grad^2 \v{E} \nonumber \\
              &=& \curl (-\pd{\v{B}}{t}) \nonumber \\
              &=& -\mu \epsilon \partial^2_t \v{E} -\mu \partial_t \v{J} + 2\alpha \mu \partial_t (\grad \theta \times \v{E}), \label{curl3}
\end{eqnarray}
where Eqs. (\ref{Maxwell1}) and (\ref{Maxwell4}) have been utilized for the third and last lines, respectively.
%
%
Assuming the constituent equation ($\v{J} = \boldsymbol{\sigma} \cdot \v{E}$), we obtain an eigenvalue equation composed of the $\v{E}$ field only,
\begin{eqnarray}
\grad (\div{\v{E}})-\grad^2 \v{E} &=& -\mu \epsilon \partial^2_t\v{E} -\mu \partial_t \boldsymbol{\sigma} \cdot \v{E} + 2\alpha \mu \partial_t \grad \theta \times \v{E}, \nonumber \label{curlcurlE} \\
\end{eqnarray}
where
\begin{equation}
\boldsymbol{\sigma} = \left(\begin{array}{ccc} (1+c_{\omega x}B_{x}^2)\sigma & \sigma_{Bz} & -\sigma_{By} \\
-\sigma_{Bz} & {(1+c_{\omega y}B_{y}^2)\sigma} & \sigma_{Bx} \\
\sigma_{B_y} & -\sigma_{Bx} & (1+c_{\omega z}B_{z}^2)\sigma \end{array}\right) . \nonumber
\end{equation}
Here, the conductivity tensor $\boldsymbol{\sigma}$ consists of both diagonal and off-diagonal components. The diagonal component $\sigma_{ii}=(1+c_{\omega i}B_{i}^2)\sigma$ is referred to as the chiral-anomaly enhanced  longitudinal magnetoconductivity, where the Drude conductivity is modified from the chiral anomaly in the Weyl metal phase \cite{WM1,WM2,WM3,WM_Reviews,Nielsen_Ninomiya}. $c_{\omega i}$ is a positive numerical constant, referred to as the chiral anomaly coefficient in the longitudinal magnetoconductivity. The off-diagonal component $\sigma_{ij} = \epsilon_{ijk} \sigma_{Bk}$ with the antisymmetric tensor $\epsilon_{ijk}$ is the conventional Hall conductivity ($\sigma_{Bk}$) driven by the $k-$th component of the external magnetic field $\v{B_{ext}}$. This conventional Hall effect results from normal electrons, which can coexist with Weyl electrons. We point out that the anomalous Hall effect has been introduced into the axion electrodynamics via $\grad{\theta}$ terms, more precisely, the last term in Eq. (\ref{curlcurlE}).

In order to solve Eq. (\ref{curlcurlE}), it is essential to deal with $\div{\v{E}}$. Here, we discuss how to find $\div{\v{E}}$ carefully since the Weyl metal physics is encoded into this quantity. More concretely, this term plays an important role in determining our normal modes inside the Weyl metal state. Our careful treatment of this term is a distinguished point, compared with the previous studies \cite{AxionEMreferee1,AxionEMreferee2,AxionEMreferee3,Axion_EM_Th_WM}. Resorting to the constituent equation of $\v{J} = \boldsymbol{\sigma} \cdot \v{E}$, we rewrite the continuity equation $\div{\v{J}} = - \pd{\rho}{t}$ as
\begin{eqnarray}
- \pd{\rho}{t} &=& \div{(\boldsymbol{\sigma} \cdot \v{E})} \nonumber \\
&\approx&\frac{\sigma_{zz}}{\epsilon} \rho + \frac{2\alpha \sigma_{zz}}{\epsilon} (\grad \theta \cdot \v{B}) ,
\label{continuity}
\end{eqnarray}
justified when $\sigma_{zz} \gg \sigma_{xy},~\sigma_{yz},~\sigma_{zx}$ and indeed for normal situations.

Considering the electromagnetic field given by $\v{B}=\v{B}_{ext}+\v{B}_{dyn}$ and $\v{B}_{dyn}=\v{B_0} e^{i(kz-\omega t)}$, one can solve Eq. (\ref{continuity}) for $\rho(t)$ as follows
\begin{eqnarray}
\rho(t) &=& -2\alpha g \sigma_{zz} \Big(\frac{\v{B^2_{ext}}}{\sigma_{zz}}+\frac{2\v{B_{ext}}\cdot \v{B_0}e^{i(kz-\omega t)}}{\sigma_{zz}-i\epsilon \omega}\nonumber \\ &&+\frac{\v{B_0}^2 e^{2(ikz-\omega t)}}{\sigma_{zz} - 2i \epsilon \omega}\Big) + C_0e^{\frac{-\sigma_{zz} t}{\epsilon}}\nonumber \\ &\approx& -2 \alpha g \Big\{B^2_{ext}+2B_{ext}B_0e^{i(kz-\omega t)} \Big(1+\frac{\omega i}{\sigma_{zz}/\epsilon} \Big) \nonumber \\ \label{rho}
&&+B_0^2e^{2i(kz-\omega t)} \Big(1+\frac{2\omega i}{\sigma_{zz}/\epsilon} \Big)\Big\} \nonumber.
\end{eqnarray}
Here, the typical metallic condition $\sigma_{zz}/\epsilon \gg \omega \gg 1$ has been used \cite{Jackson}, which allows us to keep the first order in $\frac{\omega}{\sigma_{zz}/\epsilon}$. In addition, we ignore the $e^{\frac{-\sigma t}{\epsilon}}$ term due to fast relaxation in metals \cite{Jackson}. Then, this expression can be reformulated in the following way
\begin{eqnarray}
\rho(t) &=&-2\alpha(\grad{\theta}\cdot \v{B})-4\alpha g \v{B_{ext}}\cdot \v{B_0}e^{i(kz-\omega t)} \frac{\omega i}{\sigma_{zz}/\epsilon} \nonumber \\
&&-2\alpha g B_0^2 e^{2i(kz-\omega t)}\frac{2\omega i}{\sigma_{zz}/\epsilon} . \label{Density_Harmonic_Approx}
\end{eqnarray}
The second harmonic term ($\propto e^{2i(kz-\omega t)}$) can be also negligible in the case of $|\v{B_{ext}}| \gg |\v{B_{dyn}}|$.

If the external magnetic field $\v{B_{ext}}$ is perpendicular to the oscillating field $\v{B_{dyn}}$, i.e., $\v{B_{ext}} \cdot \v{B_{dyn}}=0$, we have $\rho(t) \approx 0$ in the level of the harmonic approximation. As a result, we reach the conclusion
\begin{equation}
\div{\v{E}} = \frac{\rho}{\epsilon} + \frac{2 \alpha (\grad \theta \cdot \v{B})}{\epsilon} \approx 0.
\end{equation}
In this case which corresponds to Figs. \ref{eliptic}a and \ref{eliptic}b, the eigenmode with the longitudinal component (oscillating in $z$ direction) cannot appear due to the divergence free condition, and only transverse modes of the electromagnetic field are observable within the harmonic solution. Actually, the previous studies on this point are consistent \cite{AxionEMreferee1,AxionEMreferee2,AxionEMreferee3,Axion_EM_Th_WM}. On the other hand, if $\v{B_{ext}} \cdot \v{B_{dyn}} \neq 0$ which corresponds to Figs. \ref{eliptic}c and \ref{eliptic}d, the harmonic term is nonvanishing in the charge density [Eq. (\ref{Density_Harmonic_Approx})], resulting in
\begin{equation}
\div{\v{E}} \approx -4\alpha g \omega/\sigma_{zz} i \v{B_{ext}} \cdot \v{B_{0}}e^{i(kz-\omega t)}. \label{divE}
\end{equation}
This is rather a striking result since this condition allows the longitudinal mode inside the Weyl metal phase. We emphasize that this anomalous behavior originates from the chiral anomaly, regarded to be a characteristic feature of the axion electrodynamics. This is our point beyond the previous studies \cite{AxionEMreferee1,AxionEMreferee2,AxionEMreferee3,Axion_EM_Th_WM}. Although this aspect has been discussed very briefly in Ref. \cite{Axion_EM_Th_WM}, their eigenvectors differ from ours.

Equation (\ref{divE}) determines the first term in Eq. (\ref{curlcurlE}) as
\begin{equation}
\grad (\div{\v{E}}) \approx \frac{4\alpha g k^2}{\sigma(1+c_{\omega z}B_z^2)}\left(\begin{array}{ccc} 0 & 0 & 0 \\ 0 & 0 & 0 \\ B_y & -B_x & 0 \\ \end{array} \right) \v{E}. \label{graddiv}
\end{equation}
Then, Eq. (\ref{curlcurlE}) reads
\begin{eqnarray}
&&\left(\begin{array}{ccc}
k^2 & 0 & 0 \\
0 & k^2 & 0 \\
\frac{4\alpha gB_y}{\sigma(1+c_{\omega z}B_z^2)}k^2 & -\frac{4\alpha gB_x}{\sigma(1+c_{\omega z}B_z^2)}k^2 & k^2 \\
\end{array} \right) \v{E} \nonumber \\
&=&\mu \epsilon \omega^2\v{E} + (\mu \omega i)\boldsymbol{\sigma}\cdot \v{E} - 2\alpha g \mu \omega i \left(\begin{array}{ccc}
0 & -B_z & B_y \\
B_z & 0 & -B_x \\
-B_y & B_x & 0 \\
\end{array} \right)\v{E}. \nonumber
\end{eqnarray}
This eigenvalue equation can be reduced to a more compact form
\begin{equation}
\left(\begin{array}{ccc}
k^2-\lambda_x & id_z & -id_y \\
-id_z & k^2-\lambda_y & id_x \\
aB_y k^2 + id_y & -aB_x k^2-id_x  & k^2-\lambda_z \\
\end{array} \right) \v{E} = 0, \label{matrixEq.0}
\end{equation}
where $d_j \equiv  -\mu \omega (2\alpha g B_j + \sigma_{Bj})$, $\lambda_j \equiv \mu \epsilon \omega^2 + i\mu \omega (1+c_{\omega j}B_j^2)\sigma$, and $a \equiv \frac{4\alpha g}{\sigma(1+c_{\omega z}B_z^2)}$.

Although Eq. (\ref{matrixEq.0}) looks simple, it turns out that general expressions for eigenvalues and eigenvectors are too complex, which may not be useful for physical insight. More concretely, the eigenvalue equation is given by the third-order algebraic equation for $k^{2}$, and this equation has its general solutions according to Cardano's method \cite{Cardano_Method}. However, these general solutions would have complex numbers inside the root, and it is necessary to decompose the real and imaginary parts explicitly for such solutions in order to have physical interpretation. It turns out that this procedure is not possible, generally speaking, since the order of the algebraic equation becomes doubled for this decomposition. In this respect it is better to consider a concrete physical situation, which will be discussed from the next section. Here, we present a general solution as a reference for a specific case, given by $\sigma_{Bj} \approx 0$, $a \ll \textrm{Im}[\lambda_j]$, and $c_{wi} \approx 0$. This corresponds to the absence of the conventional Hall effect, the charge density modulation, and the positive longitudinal magnetoconductivity. We keep only the anomalous Hall effect due to the Berry curvature, which leads the matrix to be symmetric. As a result, we find three eigenvalues:
%
%
\begin{widetext}
\begin{equation}
\left\{
\begin{array}{l}
\textrm{eigenvalue 1)}\quad k_1 = k_{r1} + i k_{i1}\\
\quad k_{r1} = \omega\sqrt{\frac{\mu \epsilon}{2}}[\sqrt{(1-\frac{2\alpha g B_{ext}}{\epsilon \omega})^2+(\frac{\sigma}{\epsilon \omega})^2}+(1-\frac{2\alpha g B_{ext}}{\epsilon \omega})]^{1/2} \\
\quad k_{i1}=\omega\sqrt{\frac{\mu \epsilon}{2}}[\sqrt{(1-\frac{2\alpha g B_{ext}}{\epsilon \omega})^2+(\frac{\sigma}{\epsilon \omega})^2}-(1-\frac{2\alpha g B_{ext}}{\epsilon \omega})]^{1/2} \\
\textrm{eigenvalue 2)}\quad k_2 = k_{r2} + i k_{i2}\\
\quad k_{r2} = \omega\sqrt{\frac{\mu \epsilon}{2}}[\sqrt{(1+\frac{2\alpha g B_{ext}}{\epsilon \omega})^2+(\frac{\sigma}{\epsilon \omega})^2}+(1+\frac{2\alpha g B_{ext}}{\epsilon \omega})]^{1/2} \\
\quad k_{i2}=\omega\sqrt{\frac{\mu \epsilon}{2}}[\sqrt{(1+\frac{2\alpha g B_{ext}}{\epsilon \omega})^2+(\frac{\sigma}{\epsilon \omega})^2}-(1+\frac{2\alpha g B_{ext}}{\epsilon \omega})]^{1/2} \\
\textrm{eigenvalue 3)}\quad k_3 = k_{r3} + i k_{i3}\\
\quad k_{r3} = \omega\sqrt{\frac{\mu \epsilon}{2}}[\sqrt{1+(\frac{\sigma}{\epsilon \omega})^2}+1]^{1/2}\\
\quad k_{i3}=\omega\sqrt{\frac{\mu \epsilon}{2}}[\sqrt{1+(\frac{\sigma}{\epsilon \omega})^2}-1]^{1/2} \\
\end{array} \right. \label{eigenvaluescalarsigma}
\end{equation}
\end{widetext}
Corresponding eigenvectors are given by
\begin{equation}
\left\{
\begin{array}{lll}
\textrm{mode 1)}\quad & \left(\begin{array}{l} -B_xB_y - iB_z\sqrt{B_x^2 +B_y^2 + B_z^2} \\ B_x^2 + B_z^2 \\ -B_yB_z + iB_x\sqrt{B_x^2+B_y^2+B_z^2} \end{array}\right) \\
\textrm{mode 2)}\quad & \left(\begin{array}{l} B_xB_y - iB_z\sqrt{B_x^2 +B_y^2 + B_z^2} \\ -(B_x^2 + B_z^2) \\ B_yB_z + iB_x\sqrt{B_x^2+B_y^2+B_z^2} \end{array}\right) \\
\textrm{mode 3)}\quad & \left(\begin{array}{l} B_x \\ B_y \\ B_z \end{array}\right) \\
\end{array} \right.
\end{equation}
Here, the normalization is given by $B_{x}^{2} + B_{y}^{2} + B_{z}^{2}$.
These eigenvectors are self-consistent in the case of $\v{B_{ext}}//\hat{\v{z}}$, i.e., $B_{z} \not= 0$ and $B_{x} = B_{y} = 0$, where $\hat{\v{z}}$ is the propagating direction of light. On the other hand, they are not in the case of $\v{B_{ext}}//\hat{\v{x}}$, i.e., $B_{x} \not= 0$ and $B_{y} = B_{z} = 0$, where the existence of the longitudinal component violates the divergence free condition of the electric field. It turns out that $a = 0$ in the above approximation is not consistent for this situation. As discussed before, there appear longitudinal charge-density fluctuations, responsible for the existence of the longitudinal component of the eigenvector. In this respect $a = 0$ is just an approximation for an analytic expression. Below, we discuss full solutions for two specific directions of external magnetic fields.

\subsection{Solution in the case of $\v{B_{ext}}//\hat{\v{z}}$}

This configuration corresponds to Fig. \ref{eliptic}a. The continuity equation allows us to take $\div \v{E} \approx 0$ due to fast relaxation of electrons. As a result, Eq. (\ref{matrixEq.0}) becomes
\begin{eqnarray}
&& \left(\begin{array}{ccc} k^2-\lambda & id_z & 0 \\ -id_z & k^2 - \lambda & 0 \\ 0 & 0 & k^2 - \lambda_z \\ \end{array} \right) \v{E} = 0 , \label{matrixequation}
\end{eqnarray}
where $\lambda \equiv \mu \epsilon \omega^2 + \mu \omega i\sigma$. Obviously, the $\sigma_\text{Weyl}\equiv 2\alpha g B_z$ term in $d_z$, which is proportional to the external magnetic field, plays the role of effective Hall conductivity, and it appears only when the field is oscillating, i.e., $\omega \neq 0$. $\sigma_\text{Weyl}$ is linear in the external magnetic field whereas the conventional Hall effect or an anomalous Hall effect in a ferromagnet should be saturated for an increasing external magnetic field. This different behavior of $\sigma_\text{Weyl}$ makes characteristic magnetic-field dependencies for the propagating light in a Weyl metal state.

Considering $\v{E}=(E_x,E_y,E_z)e^{i(kz-\omega t)}$, we can find eigenvalues and eigenvectors in Eq. (\ref{matrixequation}). The eigenvalues are given by
\begin{widetext}
\begin{eqnarray}
k_{r1} &=& \omega\sqrt{\frac{\mu \epsilon}{2}}\left[\sqrt{\left(1-\frac{|1+\eta|\sigma_\text{Weyl}}{\epsilon \omega}\right)^2+(\frac{\sigma}{\epsilon \omega})^2}+\left(1-\frac{|1+\eta|\sigma_\text{Weyl}}{\epsilon \omega}\right)\right]^{1/2} , \label{kr1}\\
k_{i1} &=& \omega\sqrt{\frac{\mu \epsilon}{2}}\left[\sqrt{\left(1-\frac{|1+\eta|\sigma_\text{Weyl}}{\epsilon \omega}\right)^2+(\frac{\sigma}{\epsilon \omega})^2}-\left(1-\frac{|1+\eta|\sigma_\text{Weyl}}{\epsilon \omega}\right)\right]^{1/2} \label{ki1}
\end{eqnarray}
for the momentum $k_1 = k_{r1} + i k_{i1}$,
\begin{eqnarray}
k_{r2} &=& \omega\sqrt{\frac{\mu \epsilon}{2}}\left[\sqrt{\left(1+\frac{|1+\eta|\sigma_\text{Weyl}}{\epsilon \omega}\right)^2+(\frac{\sigma}{\epsilon \omega})^2}+\left(1+\frac{|1+\eta|\sigma_\text{Weyl}}{\epsilon \omega}\right)\right]^{1/2} \label{kr2}, \\
k_{i2} &=& \omega\sqrt{\frac{\mu \epsilon}{2}}\left[\sqrt{\left(1+\frac{|1+\eta|\sigma_\text{Weyl}}{\epsilon \omega}\right)^2+(\frac{\sigma}{\epsilon \omega})^2}-\left(1+\frac{|1+\eta|\sigma_\text{Weyl}}{\epsilon \omega}\right)\right]^{1/2} \label{ki2}
\end{eqnarray}
for the momentum $k_2 = k_{r2} + i k_{i2}$,
\end{widetext}
and
\begin{eqnarray}
k_{r3} &=& \omega\sqrt{\frac{\mu \epsilon}{2}}[\sqrt{1+(\frac{\sigma}{\epsilon \omega})^2(1+c_{\omega}B_\text{ext}^2)^2}+1]^{1/2} \label{kr3}, \\
k_{i3} &=& \omega\sqrt{\frac{\mu \epsilon}{2}}[\sqrt{1+(\frac{\sigma}{\epsilon \omega})^2(1+c_{\omega}B_\text{ext}^2)^2}-1]^{1/2} \label{ki3}
\end{eqnarray}
for the momentum $k_3 = k_{r3} + i k_{i3}$, respectively. The real part of the eigenvalues ($k_{rn}$) describes the energy of a propagating mode inside a Weyl metal (dispersion relation), and the imaginary part ($k_{in}$) gives the inverse of skin depth. Here, we introduced $\eta$ as a phenomenological parameter, defined by the ratio between the conventional Hall conductivity and the distance of the pair of Weyl points and given by
\begin{equation}
\eta \equiv \frac{\sigma_{Bz}}{\sigma_{\text{Weyl}}}.
\end{equation}
The Hall coefficient ($\sigma_{Bz}$) can be modified, depending on the sample situation in experiments. In this respect one can investigate dispersion relations as a function of $\eta$. For example, $\eta \rightarrow -1$ describes the dispersion relation of a normal metal state with $k_{rn} = \omega\sqrt{\frac{\mu \epsilon}{2}}[\sqrt{1+(\frac{\sigma}{\epsilon \omega})^2}+1]^{1/2}$ and $k_{in} = \omega\sqrt{\frac{\mu \epsilon}{2}}[\sqrt{1+(\frac{\sigma}{\epsilon \omega})^2}-1]^{1/2}$. On the other hand, $\eta \rightarrow 0$ gives rise to that of a pure Weyl metal phase without the conventional Hall effect.

The corresponding eigenvector for each of eigenvalues is given by
\begin{equation}
\begin{array}{ccc}
\textrm{ mode 1)} \left(\begin{array}{c} 1 \\ i \\ 0\\ \end{array}\right) & \textrm{ mode 2)} \left(\begin{array}{c} 1 \\ -i \\ 0 \\ \end{array}\right) & \textrm{mode 3)}
\left(\begin{array}{c} 0 \\ 0 \\ 1 \\ \end{array}\right) , \\
\label{eigenvectoreta}
\end{array}
\end{equation}
respectively, where the first two are circularly-polarized transverse modes with different chiralities and the last one is a linearly-polarized longitudinal mode. Note that for this specific configuration of $\v{B_{ext}}//\hat{z}$, the third eigenmode does not exist in the first-order harmonic approximation, which results from the divergence free condition of the electric field, i.e., $\div{\v{E}}=0 \rightarrow E_z=0$. These eigenvectors are consistent with all of the previous studies \cite{AxionEMreferee1,AxionEMreferee2,AxionEMreferee3,Axion_EM_Th_WM}. However, in general situations, for example, the case of $\v{B_{ext}}//\hat{x}$, we emphasize that there exists the contribution of charge density fluctuations according to the electromagnetic field ($\div{\v{E}} \neq 0$), so the propagating light in the Weyl metal phase should have the longitudinal component.
%
%
\begin{figure}
\centering
\includegraphics[width=9cm]{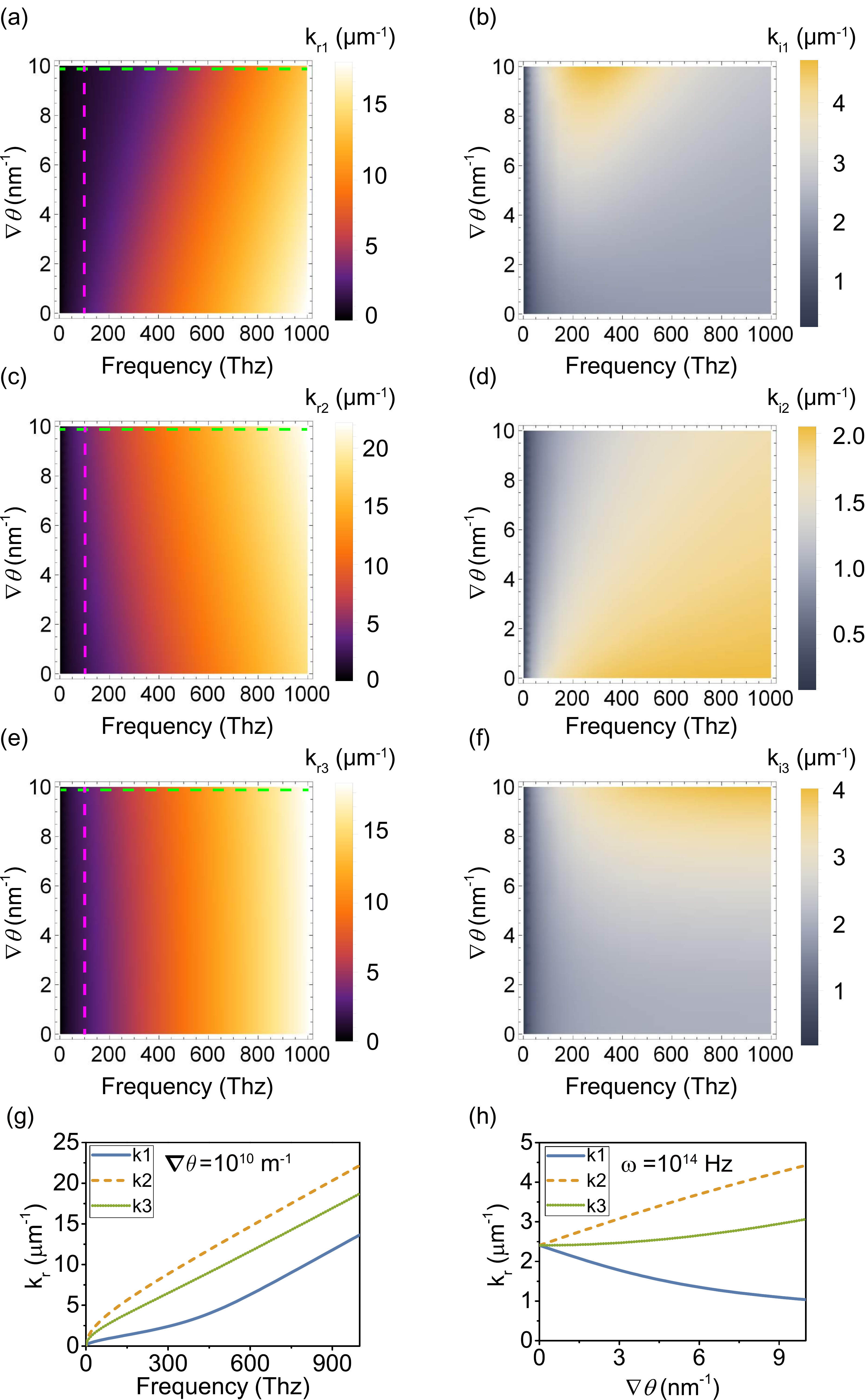}
\caption{Frequency and magnetic field dependencies for each eigenmode in a specific configuration (Fig. \ref{eliptic}a). a, c, and e. The real parts (propagating wave number) of the eigenvalue 1, 2, and 3 as a function of $\omega$ and $\grad \theta$. We recall that the $k_{3}$ mode does not exist at this configuration, but we show it as a reference. The magenta-vertical linecuts correspond to Fig. \ref{kdependence}g whereas the green-horizontal linecuts correspond to Fig. \ref{kdependence}h, respectively. b, d, and f. The imaginary parts (inverse of skin depth) of the eigenvalue 1, 2, and 3 as a function of $\omega$ and $\grad \theta$. g. Frequency dependencies of the wave number $k_r$ (dispersion relation) for each eigenmode at a finite field, $\grad \theta = 10^{10}$ m$^{-1}$. h. Magnetic field dependencies of the wave number $k_r$ for each eigenmode at a finite frequency, $\omega = 10^{14}$ Hz.}
\label{kdependence}
\end{figure}
%
%
We show frequency $\omega$ and magnetic field $\grad \theta = g \v{B_{ext}}$ dependencies of the eigenvalues with the situation $\v{B_{ext}}//\hat{\v{z}}$ in Fig. \ref{kdependence}. We set several numerical parameters to plot the eigenvalues, for example, $\sigma \sim 10^5$ $\Omega^{-1}$m$^{-1}$ and $\epsilon \sim 10\epsilon_0$, but such parameters in a real sample may be modified, thereby the appropriate frequency range in an experiment might be different from our fittings. Note that a proper frequency range of our results should be $\omega < \sigma/\epsilon$. Real (imaginary) parts of eigenvalues as a function of $\omega$ and $\grad \theta$ are described in Figs. \ref{kdependence}a, \ref{kdependence}c, and \ref{kdependence}e (\ref{kdependence}b, \ref{kdependence}d, and \ref{kdependence}f). Different features among a, c and e (b, d and f) show that propagating wave numbers (skin depth) inside a Weyl metal should be different for different modes. We recall that the $k_{3}$ mode does not exist. Here, we show it just for a reference  (Note that the dispersion of mode3 in this configuration is almost same as that of conventional metals except $c_\omega B_{ext}^2$ term). An incident light from vacuum might split into right and left circularly polarized eigenmodes. For clear presentation of splitting, a linecut graph of real $k$ at a finite magnetic field ($\grad \theta = g\v{B_{ext}}$) and at a finite frequency ($\omega$) are shown in Figs. \ref{kdependence}g and \ref{kdependence}h. Dispersion relation splits into two circularly polarized modes but their dispersions become linear at high frequency ($\omega \sim \sigma/\epsilon$) as shown in Fig. \ref{kdependence}g. On the other hand, Fig. \ref{kdependence}h shows quite different magnetic field dependencies for skin depth of each mode. When magnetic field is increasing, the differences of velocities among the modes get larger, and this splitting of a beam is the main reason of the Faraday/Kerr rotation in a Weyl metal. Note that we set $\eta = 0$ for the all plots to only see the pure Weyl metal phase, where $\eta$ is the phenomenological parameter introduced before.

\subsection{Solution in the case of $\v{B_{ext}}//\hat{\v{x}}$ \label{B//x}}

Setting the specific configuration of the external magnetic field $\v{B_{ext}}=(B_x,0,0)$ in Eq. (\ref{matrixEq.0}), which corresponds to Figs. \ref{eliptic}b, \ref{eliptic}c and \ref{eliptic}d, we find eigenvalues and eigenvectors as follows
\begin{eqnarray}
\left( \begin{array}{ccc} k^2 - \lambda_x & 0 & 0 \\ 0 & k^2-\lambda & id_x \\ 0 & -aB_x k^2 -id_x & k^2-\lambda \end{array} \right) \v{E} =0. \label{matrixEq.Bx}
\end{eqnarray}
Three eigenvalues are given by
\begin{eqnarray}
k^2_{1} &=& \lambda+\sqrt{d_x^2-i a B_x d_x k_1^2}
\end{eqnarray}
for the momentum $k_1 = k_{r1} + i k_{i1}$,
\begin{eqnarray}
k^2_{2} &=& \lambda-\sqrt{d_x^2-i a B_x d_x k_1^2}
\end{eqnarray}
for the momentum $k_2 = k_{r2} + i k_{i2}$, and
\begin{eqnarray}
k_{r3} &=& \omega\sqrt{\frac{\mu \epsilon}{2}}[\sqrt{1+(\frac{\sigma}{\epsilon \omega})^2(1+c_{\omega}B_\text{ext}^2)^2}+1]^{1/2} \label{k3rBx} \\
k_{i3} &=& \omega\sqrt{\frac{\mu \epsilon}{2}}[\sqrt{1+(\frac{\sigma}{\epsilon \omega})^2(1+c_{\omega}B_\text{ext}^2)^2}-1]^{1/2} \label{k3iBx}
\end{eqnarray}
for the momentum $k_3 = k_{r3} + i k_{i3}$. Unfortunately, only the third eigenvalue $k_3$ can be expressed in a simple analytic form, which is the same as that of the previous case. It turns out that analytic expressions for both $k_{1}$ and $k_{2}$ eigenmodes are quite complicated, which may not be useful for physical insight. Corresponding three eigenvectors are given by
\begin{eqnarray}
\begin{array}{ll}
\textrm{ mode 1)} \left(\begin{array}{c} 0 \\ \frac{1}{\sqrt{1-i\frac{aB_xk_1^2}{d_x}}} \\ i \\ \end{array}\right) & \textrm{ mode 2)} \left(\begin{array}{c} 0 \\ \frac{1}{\sqrt{1-i\frac{aB_xk_2^2}{d_x}}} \\ -i \\ \end{array}\right) \nonumber
\end{array}
\end{eqnarray}
\begin{eqnarray}
\begin{array}{l}
\textrm{mode 3)} \left(\begin{array}{c} 1 \\ 0 \\ 0\\ \end{array}\right) \label{mode3},
\end{array}
\end{eqnarray}
respectively. We point out that the complex number appears inside the root for the eigenvector. In addition, eigenmodes 1) $\&$ 2) have their longitudinal components.

In order to find their physical meaning, we take the simple approximation to ignore the $\div{\v{E}}$ contribution for the eigenvalues (not for the eigenvectors) as the zeroth-order approximation, assuming that the generated electromagnetic field from $\div{\v{E}}$ is much smaller than that from other terms. This approximation in the characteristic equation is realized by considering the condition $1 \gg |\frac{aB_xd_x}{\text{Im}[\lambda]}|=\frac{\sigma^2_{\text{Weyl}}}{\sigma^2}(1+\eta)$, and justified if the anomalous Hall conductivity due to Weyl electrons is much smaller than the Drude conductivity. Within the zeroth order approximation of $\frac{\sigma^2_{\text{Weyl}}}{\sigma^2}(1+\eta)$, we obtain the rest of eigenvalues as Eqs. (\ref{kr1}) to (\ref{ki2}), the same as those of the previous configuration. Based on these eigenvalues, the rest of the eigenvectors can be approximated, depending on the range of the parameter $\eta=\frac{\sigma_{Bx}}{\sigma_{Weyl}}$.

First, we consider $|\eta| \gg 1$ ($|\sigma_{Bx}| \gg |\sigma_{\text{Weyl}}|$), where the conventional Hall effect dominates over the anomalous Hall effect. Performing the Taylor's expansion for two small parameters of $\frac{\omega}{\sigma/\epsilon}$ and $\frac{\sigma_{\text{Weyl}}}{\sigma}$, we find the first and second eigenvectors given by
\begin{eqnarray}
\begin{array}{l}
\textrm{ mode 1)} \left(\begin{array}{c} 0 \\ \pm 1 \\ i \\ \end{array}\right) \pm i\left(\begin{array}{c} 0 \\ \frac{\sigma_{\text{Weyl}}}{\sigma}-\frac{\omega}{\sigma/\epsilon(\eta-1)} \\ 0 \\ \end{array}\right) \\
\\ \textrm{ mode 2)} \left(\begin{array}{c} 0 \\ \pm 1 \\ -i \\ \end{array}\right) \mp i\left(\begin{array}{c} 0 \\ \frac{\sigma_{\text{Weyl}}}{\sigma}+\frac{\omega}{\sigma/\epsilon(\eta-1)}  \\ 0 \\ \end{array}\right) . \label{bigeta}\\
\end{array}
\end{eqnarray}
These modes are elliptically polarized light on the $yz$ plane (see Fig. \ref{eliptic}c), which are slightly deformed from circular polarization. It is rather unexpected that two polarizations are degenerate on the $yz$ plane for a given eigenvalue, shown by both $\pm$ signs in the corresponding eigenvector. In other words, both left and right elliptically polarized light on the $yz$ plane can be an eigenvector for a given eigenvalue. This originates from non-hermiticity in the eigenvalue equation [Eq. (\ref{matrixEq.0})], given by the axion electrodynamics of $\grad \theta$.

Second, we consider $|\eta| \ll 1$ ($ |\sigma_{Bx}| \ll|\sigma_{\text{Weyl}}| $), where the anomalous Hall conductivity dominates over the conventional Hall effect. These modes are deformed in longtitudinal directions (see Fig. \ref{eliptic}d), and corresponding eigenvectors are approximated as
\begin{eqnarray}
\begin{array}{l}
\textrm{ mode 1)} \left(\begin{array}{c} 0 \\ \pm 1 \\ 1 \\ \end{array}\right) \mp i\left(\begin{array}{c} 0 \\ \frac{\sigma_{\text{Weyl}}}{\sigma} -\frac{\omega}{\sigma/\epsilon} \\ 0 \\ \end{array}\right) \\
\\ \textrm{ mode 2)} \left(\begin{array}{c} 0 \\ \pm1 \\ -1 \\ \end{array}\right) \pm i\left(\begin{array}{c} 0 \\ \frac{\sigma_{\text{Weyl}}}{\sigma} +\frac{\omega}{\sigma/\epsilon}  \\ 0 \\ \end{array}\right) .\label{smalleta}\\
\end{array}
\end{eqnarray}
Here, we multiplied $-i$ into all components in order to make the expressions ``conventional", but not essential. The major mode is linearly polarized and the minor mode is added with out of phase by $\pi/2$ for each mode. Again, we find a two-fold degeneracy with the longitudinal component.

Third, we consider $|\eta|\approx 1$ ($|\sigma_{\text{Weyl}}|\approx |\sigma_{Bx}|$), where the anomalous Hall conductivity is the same order of the conventional one. Then, the eigenmodes cannot be Taylor-expanded. Instead, we find
\begin{eqnarray}
\begin{array}{l}
\textrm{ mode 1)} \left(\begin{array}{c} 0 \\ 1/\sqrt{\frac{\sigma_{\text{Weyl}}}{\sigma}-\frac{2\omega}{\sigma/\epsilon}} \\ i\sqrt{i} \\ \end{array}\right) \\
\\ \textrm{ mode 2)} \left(\begin{array}{c} 0 \\ 1/\sqrt{\frac{\sigma_{\text{Weyl}}}{\sigma}+\frac{2\omega}{\sigma/\epsilon}} \\ -i\sqrt{i} \\ \end{array}\right) .\label{oneeta}\\
\end{array}
%
%
\end{eqnarray}
Here, we multiplied $\sqrt{i}$ into all components in order to transfer the $\sqrt{i}$ contribution of the $y$ component into the $z$ component, but not essential. If a resonance condition ($|\sigma_{\text{Weyl}}/\epsilon|\sim | 2\omega|$) is fulfilled, at least one of the denominators in the $y$ components vanishes. This means that the $y$ component dominates over the $z$ component. Then, the eigenmode 1 or 2 becomes ``consistent" to an incident beam of $\v{E_i}=(0,1,0)e^{i(kz-\omega t)}$. In this configuration, no splitting occurs and the propagating motion of the EM field is almost same as Fig. \ref{eliptic}b except the positions of $E$ and $B$ are switched. Note that $\sqrt{i}=\pm\frac{i+1}{\sqrt{2}}$ has both $\pm$ signs. As a result, we have a two-fold degeneracy with the longitudinal component again at both eigenvalues of $k_{1}$ and $k_{2}$.

In this configuration ($\v{B_{ext}}//\hat{\bm{x}}$) the mode 1 and mode 2 have different polarization directions, given by different phases in their longitudinal oscillations. However, we emphasize that these solutions are completely consistent with $\div{\v{B}}=0$ and all the other Maxwell's equations. In other words, only the electric field is oscillating to be described with 3 degrees of freedom, whereas the magnetic field is oscillating with 2 transverse degrees of freedom. See Figs. \ref{eliptic}c and \ref{eliptic}d. Of course, all eigenvectors satisfy the modified Maxwell's equations [Eqs. (\ref{Maxwell1}) $\sim$ (\ref{Maxwell4})].

To visualize this point clearly, we present Fig. \ref{eliptic}. Figure \ref{eliptic}a corresponds to the case that the external magnetic field is applied into the $\hat{\v{z}}-$direction, where the mode 3 does not exist due to the divergence free condition ($\nabla \cdot \bold{E} = 0$ and $\nabla \cdot \bold{B} = 0$). Both electric and magnetic fields have two transverse degrees of freedom and the longitudinal oscillation in the $\hat{\v{z}}-$direction is forbidden. When the external magnetic field is applied into the $\hat{\v{x}}-$direction (Figs. \ref{eliptic}b, \ref{eliptic}c, and \ref{eliptic}d), the longitudinal oscillation of the electric field is allowed due to the non-zero divergence, i.e., $\nabla \cdot \bold{E} = \frac{\rho}{\epsilon} + 2 \alpha g \nabla \theta \cdot \bold{B} \neq 0$, which is one essential modification in the axion electrodynamics. Here, the mode 3 is just a conventional electromagnetic field configuration with only transverse oscillating components ($\vec{\v{E}}//\hat{\bm{x}}$ and $\vec{\v{B}}//\hat{\bm{y}}$) as shown in Fig. \ref{eliptic}b. On the other hand, the mode 1 and mode 2 have longitudinally oscillating components. For the $\eta \gg 1$ case (Fig. \ref{eliptic}c), two elliptically polarized electric fields are possible for the mode 1 and mode 2. The oscillation of the $B$ field is in the $\hat{\v{x}}-$direction for both modes. For the $\eta \ll 1$ case (Fig. \ref{eliptic}d), two linearly polarized electric fields are possible for each mode. The oscillation of the $B$ field is in the $\hat{\v{x}}-$direction again for either mode. Consequently, there can exist three eigenvectors with two transverse degrees of freedom in the $B$ field oscillation. 

\section{Transmission and reflection of light}

\begin{figure*}
\centering
\includegraphics[width=18cm]{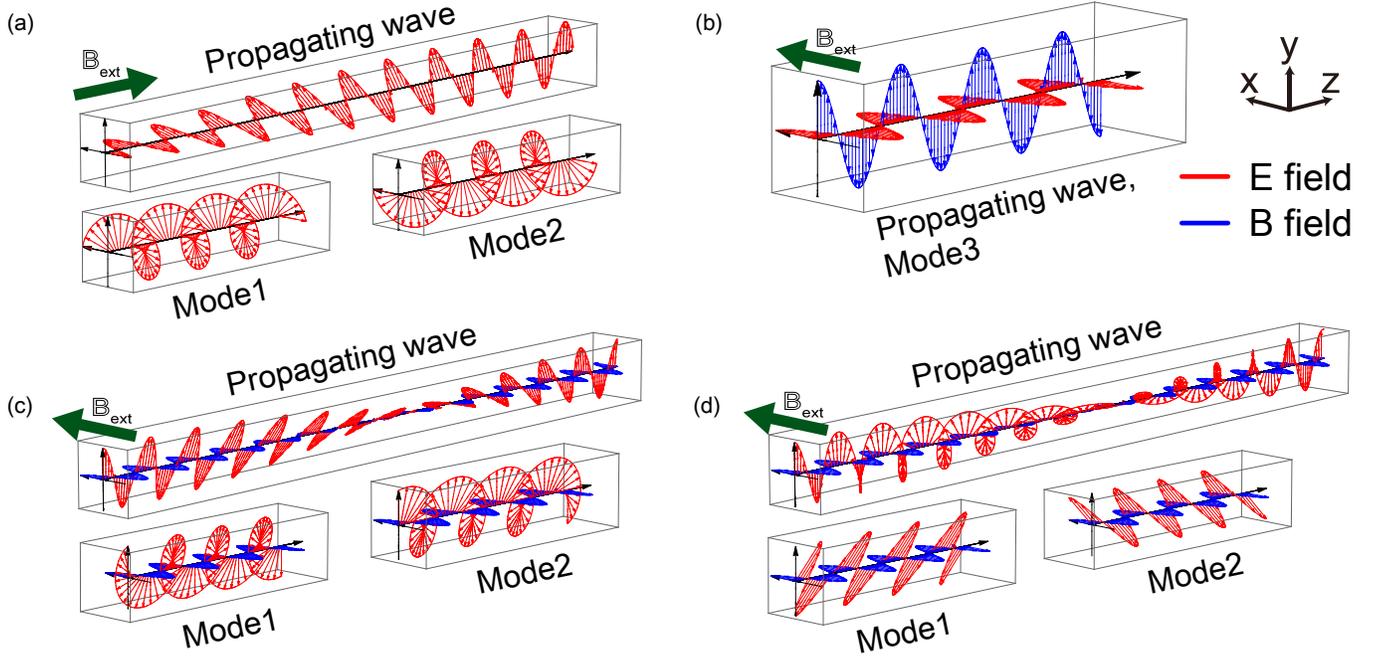}
\caption{Polarizations and rotations of the electromagnetic field inside the Weyl metal phase. a. A deformation and the Faraday-Kerr rotation when $\v{B_\text{ext}}//\bm{\hat{z}}$ and $\v{E_i}//\bm{\hat{x}}$: The rotation angle is on the $xy$ plane. The propagating wave is a consequence of the linear combination of two circularly polarized mode 1 and 2. There is no mode 3 when $\v{B_\text{ext}}//\bm{\hat{z}}$ as discussed in the main text. b. No deformation occurs when $\v{B_\text{ext}}// \bm{\hat{x}}$ and $\v{E_i}// \bm{\hat{x}}$. The incident beam and propagating wave is consistent with the mode 3. In this configuration, no deformation occurs but reflectivity changes as a function of the external magnetic field. It represents a unique property of Weyl metals as described in section \ref{reflectivity change}. c. A deformation of the transmitted light when $\v{B_\text{ext}}// \bm{\hat{x}}$ and $\v{E_i}// \bm{\hat{y}}$ with $\eta \gg 1$: The oscillating direction of the propagating electric field is rotating on the $yz$ plane. The propagating wave is a linear combination of elliptically polarized two modes, where the polarization axis of the mode is not the propagating direction (not the $\bm{\hat{z}}$ axis), but the $\grad \theta$ direction (the $\bm{\hat{x}}$ axis). There exists a beat in the propagating wave due to different group velocities of the two modes. d. A deformation of the transmitted light when $\v{B_\text{ext}}// \bm{\hat{x}}$ and $\v{E_i}// \bm{\hat{y}}$ with $\eta \ll 1$: The oscillating direction of the propagating electric field is rotating on the $yz$ plane. The propagating wave is a linear combination of linearly polarized two modes whose oscillating directions are approximately (0,1,1) and (0,1,-1). A beat also exists in the propagation wave due to different group velocities of the two modes. We point out that for the transmitted/reflected beam of Figs. \ref{eliptic}c and \ref{eliptic}d in a vacuum, the longitudinal component can exist only in the near-field region because it radiates as a short-dipole-antenna field near the surface, completely consistent with boundary conditions.}
\label{eliptic}
\end{figure*}

We are ready to calculate Faraday/Kerr rotation angles and transmission/reflection coefficients for specific situations of the alignment of $\v{B_\text{ext}}$ and the polarization of the incident beam. Here, we focus on three cases based on the solutions in the previous section: one with $\v{B_\text{ext}}//\bm{\hat{z}}$ and the other two with $\v{B_\text{ext}}//\bm{\hat{x}}$, where light always propagates along the $\bm{\hat{z}}-$direction (Figs. \ref{eliptic}a $\sim$ \ref{eliptic}d).

\begin{figure*}
\centering
\includegraphics[width=18cm]{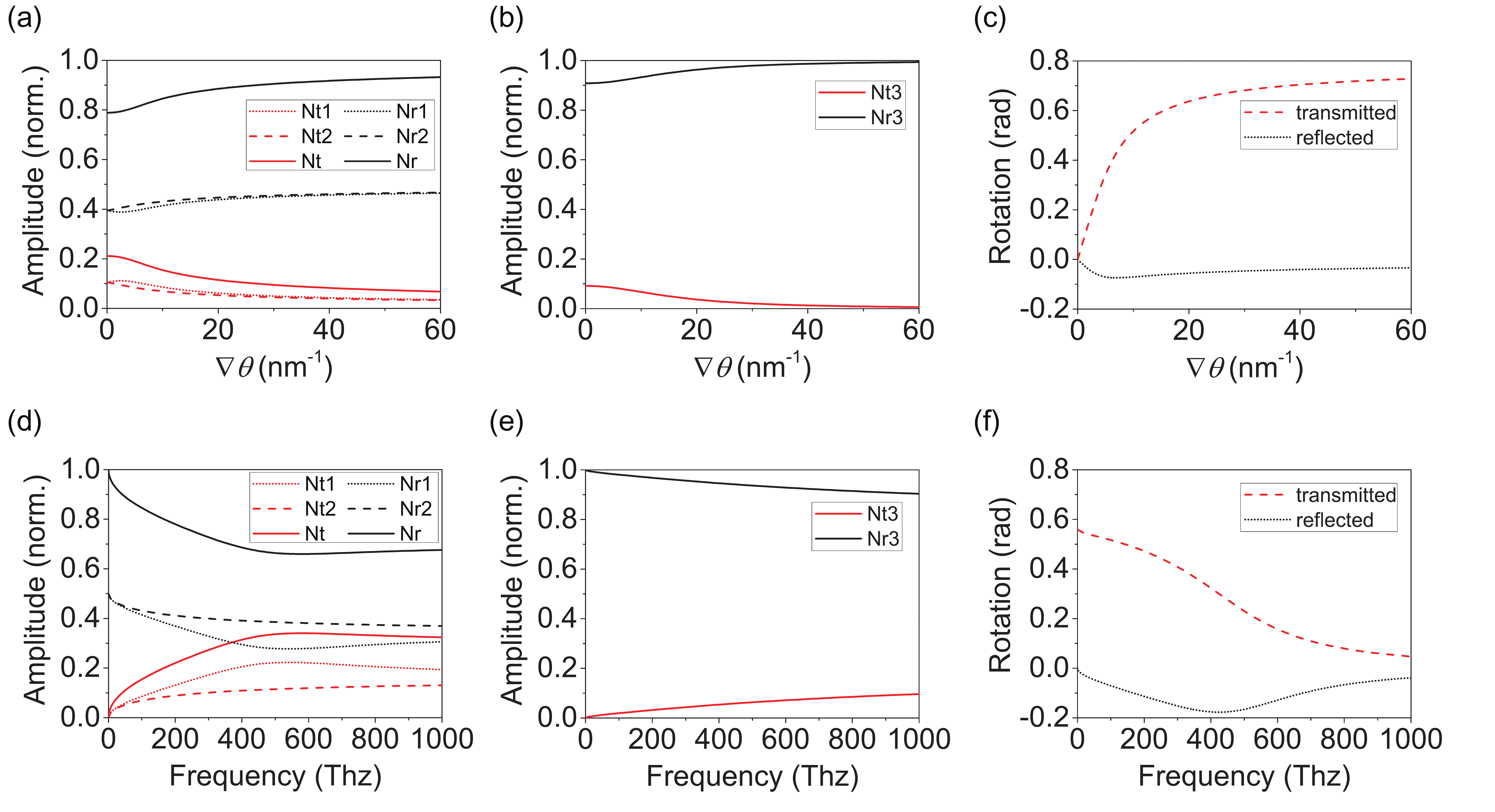}
\caption{Transmission-reflection and Faraday-Kerr rotation angles as a function of external magnetic fields (at a given frequency $\omega=10^{14}$ Hz) and frequencies (at a given magnetic field $\grad \theta=10^{10}$ m$^{-1}$). a $\&$ d. Amplitudes of transmission/reflection (red lines/black lines) depending on the applied magnetic field \& frequency. Figure. \ref{eliptic}a shows the experimental situation. $N_{i} = N_{i1} + N_{i2}$ is a total transmission/reflection amplitude which is sum of the mode 1 and 2 ($i = t$ or $r$). $N_{ij}$ ($j= 1$ or $2$) is the amplitude of each transmitted/reflected eigenmode. b $\&$ e. Amplitudes of transmission/reflection of the third eigenmode depending on the applied magnetic field \& frequency. Figure. \ref{eliptic}b shows the experimental situation. $N_{t3}$ and $N_{r3}$ are the amplitude of the transmitted and reflected eigenmode. All amplitudes of a, b, d, and e are normalized by the incident beam. c \& f. Faraday/Kerr rotation angles as a function of the applied magnetic field \& frequency. The Faraday angle is an increasing function of the applied magnetic field at a given frequency and a decreasing function of frequency at a given external magnetic field, whereas the Kerr angle shows non-monotonic behaviors in both cases.}
\label{Maximum_amplitude_Rotation_angle_for_a_linear_incident_beam}
\end{figure*}

\subsection{$\v{B_{\text{ext}}}$ // \v{\hat{z}}} \label{B//z}

First, we consider the case when $\v{B_\text{ext}}$ is aligned with the propagation of light, as shown in Fig. \ref{eliptic}a. Inside and outside of the Weyl metal, the electric and magnetic fields are given by
\begin{equation}
\textrm{$\v{E}$ field} \left \{
\begin{array}{lll}
\v{E_{in}} &=& \v{E}_{Ti}e^{ik_{Ti} z -\omega t}   \\
\v{E_{out}} &=& \v{E}_Ie^{ik_0 z -\omega t} + \v{E}_{Ri}e^{ik_{Ri} z -\omega t} \\
\end{array} \right. \nonumber
\end{equation}
and
\begin{equation}
\textrm{$\v{B}$ field} \left \{
\begin{array}{lll}
\v{B_{in}} &=& \v{B}_{Ti}e^{ik_{Ti} z -\omega t}   \\
\v{B_{out}} &=& \v{B}_Ie^{ik_0 z -\omega t} + \v{B}_{Ri}e^{ik_{Ri} z -\omega t} \\
\end{array} \right. , \nonumber
\end{equation}
respectively.  $\v{E_{in}}$ ($\v{B_{in}}$) and $\v{E_{out}}$ ($\v{B_{out}}$) correspond to the electric (magnetic) field inside and outside of the Weyl metal, where $\v{E}_I$ is the amplitude of the incident beam, $k_0$ is the wave number of the incident beam, $k_{Ti}$ ($k_{Ri}$) is the wave number of the transmitted (reflected) beam, and $\v{E}_{Ti}$ ($\v{E}_{Ri})$/$\v{B}_{Ti}$ ($\v{B}_{Ri}$) is the transmitted (reflected) amplitude of the electric (magnetic) field with an $i-$th mode. Then, amplitudes of these electric and magnetic fields at the interface ($z=0$) are given by
\begin{equation}
\textrm{$\v{E}$ field} \left \{
\begin{array}{lll}
\textrm{Incident beam}&\v{E}_I=& E_I^x \hat{x} + E_I^y \hat{y} \\
\textrm{Reflected beam}&\v{E}_{Ri}=& E_{Ri}^x \hat{x} + E_{Ri}^y \hat{y} + E_{Ri}^z \hat{z}\\
\textrm{Transmitted beam}&\v{E}_{Ti}=& E_{Ti}^x \hat{x} + E_{Ti}^y \hat{y} + E_{Ti}^z \hat{z} \\
\end{array} \right. \nonumber
\end{equation}
and
\begin{equation}
\textrm{$\v{B}$ field} \left \{
\begin{array}{lll}
\textrm{Incident beam}&\v{B}_I=& B_I^x \hat{x} + B_I^y \hat{y} \\
\textrm{Reflected beam}&\v{B}_{Ri}=& B_{Ri}^x \hat{x} + B_{Ri}^y \hat{y} \\
\textrm{Transmitted beam}&\v{B}_{Ti}=& B_{Ti}^x \hat{x} + B_{Ti}^y \hat{y} \\
\end{array} \right. , \nonumber
\end{equation}
respectively. Note that only $E_T$ and $E_R$ can have an oscillating $z-$component (parallel to the propagating direction) because the incident beam is propagating in vacuum without a source term and Eq. (\ref{Maxwell3}) does not allow the magnetic field to have a $z-$component. With the divergence-free condition and Eq. (\ref{Maxwell3}), the boundary conditions at the interface for each mode give rise to the following eight equations,
\begin{equation}\label{Boundary_Condition_1}
\begin{array}{cccc}
1)&\quad E_I^x+E_{Ri}^x &=& E_{Ti}^x \\
2)&\quad E_I^y+E_{Ri}^y &=& E_{Ti}^y \\
3)&\quad B_I^x+B_{Ri}^x &=& B_{Ti}^x \\
4)&\quad B_I^y+B_{Ri}^y &=& B_{Ti}^y \\
5)&\quad B_{Ti}^x &=& -\frac{k}{\omega}E_{Ti}^y \\
6)&\quad B_{Ti}^y &=& \frac{k}{\omega}E_{Ti}^x \\
7)&\quad B_{Ri}^x &=& \frac{1}{c}E_{Ri}^y \\
8)&\quad B_{Ri}^y &=& -\frac{1}{c}E_{Ri}^x \\
\end{array} .
\end{equation}
Here, $E_I^x$, $E_I^y$, $B_I^x$, and $B_I^y$ serve as initial conditions. These eight equations fix the eight unknown variables completely.

Applying these boundary conditions to each $i-$th mode, we obtain the transmission/reflection coefficient for each eigenmode
\begin{eqnarray}
\textrm{Transmission coefficient : } T_i &=& \frac{2}{1+\beta_i} \label{Tscalar} \\
\textrm{Reflection coefficient : } R_i &=& \frac{1-\beta_i}{1+\beta_i} \label{Rscalar}
\end{eqnarray}
with $\beta_i = k_i/k_0$, where $k_i$ is the $i-$th eigenvalue. Now we have the transmission/reflection coefficient for each mode. We can decompose an incident beam $(a,b,c)$ in a proper linear combination of the three eigenmodes and analyze the total transmission in each basis as following,
\begin{eqnarray} \label{decompose}
\v{T} \left(\begin{array}{l} a \\ b \\ c \end{array}\right) &=& T_1\frac{\left(a-bi\right)}{2} \left(\begin{array}{l} 1 \\ i \\ 0 \end{array}\right) +
T_2\frac{\left(a+bi\right)}{2} \left(\begin{array}{l} 1 \\ -i \\ 0 \end{array}\right) \nonumber \\ &+& T_3 c\left(\begin{array}{l} 0 \\ 0 \\ 1 \end{array}\right) .
%
%
\end{eqnarray}

Putting each $T_i$ given by Eq. (\ref{Tscalar}) to Eq. (\ref{decompose}), we find the total transmission matrix ($\v{T}$) as
\begin{eqnarray}
\v{T} = \left(\begin{array}{ccc} \frac{1}{1+\beta_1} + \frac{1}{1+\beta_2} & -i(\frac{1}{1+\beta_1} - \frac{1}{1+\beta_2}) & 0 \\
i(\frac{1}{1+\beta_1} - \frac{1}{1+\beta_2}) & \frac{1}{1+\beta_1} + \frac{1}{1+\beta_2} & 0 \\
0 &0 & \frac{2}{1+\beta_3}  \end{array}\right) . \label{BzTgeneral}
\end{eqnarray}
Similarly, the total reflection matrix ($\v{R}$) is given by
\begin{eqnarray}
\v{R} = \frac{1}{2}\left(\begin{array}{ccc} \frac{1-\beta_1}{1+\beta_1} + \frac{1-\beta_2}{1+\beta_2} & -i(\frac{1-\beta_1}{1+\beta_1} - \frac{1-\beta_2}{1+\beta_2}) & 0 \\
i(\frac{1-\beta_1}{1+\beta_1} - \frac{1-\beta_2}{1+\beta_2}) & \frac{1-\beta_1}{1+\beta_1} + \frac{1-\beta_2}{1+\beta_2} & 0 \\
0 &0 & \frac{2(1-\beta_3)}{1+\beta_3}  \end{array}\right) . \label{BzRgeneral} \nonumber \\
\end{eqnarray}

Shining a linearly polarized beam into the Weyl metal state with $\v{E_i}=(E_0 e^{i(kz-\omega t)},0,0)$, we obtain
\begin{eqnarray}
\v{E_t} &=& \left(\frac{2}{1+\beta_1}+\frac{2}{1+\beta_2}, i(\frac{2}{1+\beta_1}-\frac{2}{1+\beta_2}), 0 \right) \frac{E_0}{2}e^{i(-\omega t)} \label{BzT} \nonumber \\ \\
\v{E_r} &=& \left(\frac{1-\beta_1}{1+\beta_1}+\frac{1-\beta_2}{1+\beta_2}, i(\frac{1-\beta_1}{1+\beta_1}-\frac{1-\beta_2}{1+\beta_2}), 0 \right) \frac{E_0}{2}e^{i(-\omega t)} \label{BzR} \nonumber \\
\end{eqnarray}
at the boundary of $z = 0$. Introducing $N_{ti}e^{i\phi_{ti}} \equiv T_i = \frac{2}{1+\beta_i}$ and $N_{ri}e^{i\phi_{ri}}\equiv R_i = \frac{1-\beta_i}{1+\beta_i}$, where imaginary parts can be easily eliminated, we find their real parts
\begin{eqnarray}
Re(\v{E_t}) &=& \frac{E_0}{2}\left( \begin{array}{c}N_{t1}\cos{(\omega t - \phi_{t1})}+N_{t2}\cos{(\omega t - \phi_{t2})}
\\ N_{t1}\sin{(\omega t - \phi_{t1})}-N_{t2}\sin{(\omega t - \phi_{t2})}
\\ 0 \end{array} \right) \label{ReEtBz} \nonumber \\ \\
Re(\v{E_r}) &=& \frac{E_0}{2}\left( \begin{array}{c}N_{r1}\cos{(\omega t - \phi_{r1})}+N_{r2}\cos{(\omega t - \phi_{r2})}
\\ N_{r1}\sin{(\omega t - \phi_{r1})}-N_{r2}\sin{(\omega t - \phi_{r2})}
\\ 0 \end{array} \right) . \label{ReErBz} \nonumber \\
\end{eqnarray}
Instead of linearly polarized incident beams, we get elliptically polarized transmitted (reflected) beams. The maximum amplitude is modified by the factor of $\frac{(N_{t1}+N_{t2})}{2}$ ($\frac{(N_{r1}+N_{r2})}{2}$), whereas the major axis is rotated by $\phi = \frac{\phi_{t2}-\phi_{t1}}{2}$ ($\frac{\phi_{r2}-\phi_{r1}}{2}$) from that of the incident beam. Here, we introduced
\begin{eqnarray}
\phi_{tj} &\equiv& \arctan{\left(\frac{-\text{Im}[\beta_{j}]}{1+\text{Re}[\beta_{j}]}\right)} \\
\phi_{rj} &\equiv& \arctan{\left(\frac{-2 \text{Im}[\beta_{j}]}{1-\text{Re}[\beta_{j}]^2-\text{Im}[\beta_{j}]^2}\right)} \\
N_{tj} &\equiv& \frac{2}{\sqrt{1+\text{Re}[\beta_{j}]^2+\text{Im}[\beta_{j}]^2}} \\
N_{rj} &\equiv& \frac{\sqrt{(1-\text{Re}[\beta_{j}]^2-\text{Im}[\beta_{j}]^2)^2+4\text{Im}[\beta_{j}]^2}}{(1+\text{Re}[\beta_{j}]^2)^2+\text{Im}[\beta_{j}]^2}.
\end{eqnarray}

This elliptical shape of the electric field and the rotation of the major axis at the interface originates from the splitting of light into eigenmodes at the Weyl metal. The incident beam $(1,0,0)$ is not an eigenmode inside the Weyl metal state, thereby it is decomposed into the mode $1$ and $2$ ($(1,i,0)$ and $(1,-i,0)$) at the interface. These (right- and left- handed) circularly polarized beams have different transmission (reflection) coefficients $T_i$ ($R_i$) in the Weyl metal phase. Therefore, we observe differences of their phases and amplitudes. Non-vanishing components perpendicular to the incident beam arise from the inhomogeneous $\theta-$term which makes the Faraday/Kerr rotation with an elliptical shape of the beam.

Not only at the interface but also inside the Weyl metal, the elliptical deformation and its rotation occur. When the light propagates during a distance $D$, we get a transmitted electric field, just modified by the exponential argument from $i(-\omega t)$ to $i(kD-\omega t)$. Extracting out only the real part, we obtain
\begin{eqnarray}
Re(\v{E_D}) &=& \frac{E_0}{2}\left( \begin{array}{c}N_{D1}\cos{(\omega t - \phi_{D1})}+N_{D2}\cos{(\omega t - \phi_{D2})}
\\ N_{D1}\sin{(\omega t - \phi_{D1})}-N_{D2}\sin{(\omega t - \phi_{D2})}
\\ 0 \end{array} \right) \nonumber
\end{eqnarray}
where $N_{Dj} = N_{tj}e^{-\text{Im}[k_{j}]D}$ and $\phi_{Dj} \equiv \phi_{tj}+\text{Re}[k_{j}]D$. The amplitude of $N_{Dj}$ is reduced by a factor of $e^{-\text{Im}[k_{j}]D}$ from $N_{tj}$. The phase of $\phi_{Dj}$ is shifted more by $\text{Re}[k_{j}]D$ from $\phi_{tj}$, so the total rotation angle of the major axis is $\phi=\frac{\phi_{t2}-\phi_{t1}}{2}+\frac{k_{r2}-k_{r1}}{2}D$. Here, the first term ($\frac{\phi_{t2}-\phi_{t1}}{2}$) corresponds to the rotation at the interface, and the second term ($\frac{k_{r2}-k_{r1}}{2}D$) is the rotation proportional to the propagating length inside the Weyl metal, just similar to the conventional Faraday rotation.

\subsection{$\v{B_\text{ext}} // \v{\hat{x}}$ $\&$ $\v{E_i} // \v{\hat{x}}$ \label{reflectivity change}}

When the external magnetic field is not parallel with the propagating direction of light, the divergence of the electric field could be non-vanishing, given by $\div{\v{E}} \approx -\frac{4\alpha g \omega}{\sigma(1+c_{\omega z} B_z^2)}i \v{B_{ext}}\cdot \v{B_0}e^{i(kz-\omega t)} \neq 0$. Then, the polarization direction of $\v{E_i}$ and $\v{B_i}$ is important. We are considering the incident beam as $\v{E_i}=(E_0 e^{i(kz-\omega t)},0,0)$ (see Fig. \ref{eliptic}b). In this case the incident beam is ``consistent" with the mode $3$ in Eq. (\ref{mode3}). Boundary conditions can be applied in the same as those of the normal metal state. As a result, the transmission/reflection coefficients are given by the scalar form in Eqs. (\ref{Tscalar}) and (\ref{Rscalar}). There is no Faraday/Kerr rotation, and the eigenvalue $k_3$ (see Eqs. (\ref{k3rBx}) and (\ref{k3iBx})) resembles that of a normal metal except for the $B^{2}-$enhancement factor. However, this $B^{2}$ magnetic-field dependence, which can be found from the topologically generalized Boltzmann transport theory \cite{CME3,CME4,CME5,CME6,CME7,Boltzmann_Chiral_Anomaly1,Boltzmann_Chiral_Anomaly2,Boltzmann_Chiral_Anomaly3,Boltzmann_Chiral_Anomaly4,Boltzmann_Chiral_Anomaly5,Boltzmann_Chiral_Anomaly6,Boltzmann_Chiral_Anomaly7,
Boltzmann_Chiral_Anomaly8,NLMR_First_Exp,NLMR_Followup_VIII,Disordered_Weyl_Metal2}, represents the unique property of the Weyl metal phase on this configuration ($\v{E_i}//\v{B_{ext}}$); Weyl metals become more ``reflective" with an increasing external magnetic field. Considering the configuration of Fig. \ref{eliptic}b, the reflectivity of the Weyl metal is enhanced as a function of the applied magnetic field due to the longitudinal magnetoconductivity enhanced by the $B^2$ factor as shown in Fig. \ref{Maximum_amplitude_Rotation_angle_for_a_linear_incident_beam}b. This originates from the existence of a perfect metallic channel, referred to as the chiral anomaly \cite{WM1,WM2,WM3,WM_Reviews,Nielsen_Ninomiya}.

\subsection{$\v{B_\text{ext}}//\v{\hat{x}}$ $\&$ $\v{E_i}//\v{\hat{y}}$}

Finally, we consider the case of Figs. \ref{eliptic}c and \ref{eliptic}d with an incident beam $\v{E_i}=(0,E_0 e^{i(kz-\omega t)},0)$. The incident beam is not an eigenmode in the Weyl metal state, thereby it should be decomposed into the mode $1$ and $2$ at the interface. Accordingly, we consider the following boundary conditions, which differ from those of Fig. \ref{eliptic}a,
\begin{displaymath}
\begin{array}{ccccc}
1)&\quad E_I^y+E_R^y &=& E_T^y& \\
2)&\quad \epsilon_0(E_I^z+E_R^z) - \epsilon E_T^z &=& \rho_{s} & \\
3)&\quad (B_I^x+B_R^x)/\mu_0 -B_T^x/\mu &=& J_s& \\
4)&\quad B_T^x &=& -\frac{k}{\omega}E_T^y& \\
5)&\quad (-1)^{j-1}(i\sqrt{1-i\frac{aB_xk_j^2}{d_x}})E_T^y&=& E_T^z& (\text{$j$=1 or 2})\\
6)&\quad (-1)^{j-1}(i\sqrt{1-i\frac{aB_xk_j^2}{d_x}})E_R^y&=& E_R^z& (\text{$j$=1 or 2}),
\end{array}
\end{displaymath}
where $\rho_s$ and $J_s$ are charge and current density at the interface, respectively. The surface charge density is given by Eq. (\ref{rho}), involved with the axion electrodynamics. One can obtain the surface current density via the constituent relation, given by the surface conductivity $\sigma_{s}$. For the surface current density, we would like to refer detailed discussions to Ref. \cite{Axion_EM_Th_WM}. All the parameters of $a$, $B_x$, $k_j$, and $d_x$ are already introduced in section \ref{Axionem}, where $j=1$ or $2$ corresponds to the mode 1 or 2. Both electric and magnetic fields at the boundary should be considered as
\begin{equation}
\textrm{$\v{E}$ field}\left\{
\begin{array}{lll}
\textrm{Incident beam}&=& E_I^y \hat{y} + E_I^z \hat{z} \\
\textrm{Reflected beam}&=& E_R^y \hat{y} + E_R^z \hat{z} \\
\textrm{Transmitted beam}&=& E_T^y \hat{y} + E_T^z \hat{z} \\
\end{array} \right. \nonumber
\end{equation}
and
\begin{equation}
\textrm{$\v{B}$ field}\left\{
\begin{array}{lll}
\textrm{Incident beam}&=& B_I^x \hat{x} \\
\textrm{Reflected beam}&=& B_R^x \hat{x}\\
\textrm{Transmitted beam}&=& B_T^x \hat{x}\\
\end{array} \right. , \nonumber
\end{equation}
respectively.

As discussed in section \ref{generalsol}, the explicit form of eigenvalues and eigenvectors are rather complicated to use. For physical insight, we consider the $|\eta|\gg1$ condition, following the same strategy of section \ref{B//z}, where the conventional Hall effect is dominant. Considering major eigenmodes in Eq. (\ref{bigeta}), we find the electric fields from the boundary conditions
\begin{eqnarray}
Re(\v{E_t}) &=& \frac{E_0}{2}\left( \begin{array}{c} 0
\\ N_{t1}\cos{(\omega t - \phi_{t1})}+N_{t2}\cos{(\omega t - \phi_{t2})}
\\ N_{t1}\sin{(\omega t - \phi_{t1})}-N_{t2}\sin{(\omega t - \phi_{t2})}
\end{array} \right) \nonumber \label{ReEtEy} \\ \\
Re(\v{E_r}) &=& \frac{E_0}{2}\left( \begin{array}{c} 0
\\ N_{r1}\cos{(\omega t - \phi_{r1})}+N_{r2}\cos{(\omega t - \phi_{r2})}
\\ N_{r1}\sin{(\omega t - \phi_{r1})}-N_{r2}\sin{(\omega t - \phi_{r2})}
\end{array} \right) \nonumber . \label{ReErEy} \\
\end{eqnarray}
We note that the zeroth order approximation for $|\eta|$ results in the same eigenmodes and eigenvalues with parameters $N_{ij}$ and $\phi_{ij}$ as the case of section \ref{B//z}. These electric fields are most dominant but there could exist minor corrections as described in Eqs. (\ref{bigeta}) to (\ref{oneeta}). The only difference compared to those of the case of section \ref{B//z} within this approximation is in the rotation direction of the major axis. In particular, a longitudinal component turns out to appear in both transmission ($Re(\v{E_t})$) and reflection ($Re(\v{E_t})$). See Eqs. (\ref{ReEtEy}) and (\ref{ReErEy}). In order to match boundary conditions, the longitudinal component is indispensable for the transmitted/reflected light ($Re(\v{E_t})$/$Re(\v{E_r})$) in vacuum. Remember that the oscillating $\hat{z}$ component is allowed if and only if there exist charge oscillations due to the divergence term. We recall Eqs. (\ref{continuity}) to (\ref{divE}). It is natural to expect that the longitudinal component $(0,0,1)$ of the transmitted or reflected beam in vacuum should be observed only in the near-field region ($r \ll \lambda_0$) from the interface because the charge oscillation can exist only inside the Weyl metal. Here, $\lambda_0$ is the wavelength of light in the vacuum. On the other hand, the $(0,1,0)$ component can propagate without radiation in vacuum just as the conventional electromagnetic wave.

In an experimental situation with a detector located at a far-field region ($r \gg \lambda_0$), the $\bm{\hat{z}}-$component of the transmitted/reflected beam ($E_{rz}$) should be observed as the radiation pattern of a short dipole antenna due to the charge oscillation at the interface ($E_{rz}\approx \frac{\sigma_s KE_0(N_{r1}+N_{r2})}{2\omega r}\sin{\frac{\phi_{r2}-\phi_{r1}}{2}}\sin{\theta_z}\cos{(kz-\omega t-\frac{\phi_{2r}-\phi_{1r}}{2}})$). Here, $E_0$ is an amplitude of the incident beam, $\theta_z$ is a polar angle, $r$ is a distance from the center of the shined area, $K$ is a geometrical factor of shined area, and $\sigma_s$ is the conductivity at the surface \cite{Jackson}. This dipole antenna solution is from the oscillating charge accumulation at the surface. Suppose a periodic boundary of the sample \cite{Jackson}. Then, the continuity equation should be satisfied as $\div{\v{J}}=\nabla \cdot \sigma_s\v{E}=-\pd{\rho}{t}$ at the interface, where $\sigma_s$ is a surface conductivity. Applying the Gauss theorem to the whole sample surface, we get $\int E_r^z dS = \frac{i\omega}{\sigma_s}\int \rho dV$ $\longrightarrow$ $E_r^z = \frac{i\omega}{\sigma_s} \int \rho dz$. Considering a surface charge density $n_q \equiv \int \rho dz$ given by an oscillating surface charge density $Re(n_q) = \frac{\sigma E_0}{2\omega}\left(N_{r1}\sin{(\omega t-\phi_{r1})}-N_{r2}\sin{(\omega t - \phi_{r2})}\right)$, we obtain the radiation electric field with the longitudinal component. In other words, we may consider this oscillating surface charge density as a source of the radiation of the $\bm{\hat{z}}-$directional field in the vacuum space.

We emphasize that the propagating wave shows beating inside the Weyl metal phase (see Figs. \ref{eliptic}c and \ref{eliptic}d). The beating phenomenon originates from splitting of the incident beam into two different modes with different group velocities. One may point out that the Faraday rotation in Fig. \ref{eliptic}a also occurs in ferromagnetic materials. In this case only the transverse components of the propagating wave are existing and rotating. However, the longitudinal component of the propagating electric field appears to show beating in Weyl metals, regarded to be the manifestation of longitudinal charge density wave fluctuations and resulting from the axion electrodynamics.

We summarize all these features in Fig. \ref{Maximum_amplitude_Rotation_angle_for_a_linear_incident_beam}, showing transmission/reflection coefficients and Faraday/Kerr rotation angles as a function of both external magnetic fields and light frequencies in the experimental setup of Fig. \ref{eliptic}a and Fig. \ref{eliptic}b. Figures \ref{Maximum_amplitude_Rotation_angle_for_a_linear_incident_beam}a (b) and \ref{Maximum_amplitude_Rotation_angle_for_a_linear_incident_beam}d (e) show maximum amplitudes of transmission/reflection coefficients at the situation of Fig. \ref{eliptic}a (Fig. \ref{eliptic}b). $N_{t} = N_{t1} + N_{t2}$ ($N_{r} = N_{r1} + N_{r2}$) is a total maximum transmission (reflection) amplitude, where $N_{ti}$ ($N_{ri}$) with $i = 1 ~ \& ~ 2$ is that of the transmitted (reflected) eigenmode. The external magnetic-field dependence of reflectivity enhancement at the configuration of Fig. \ref{eliptic}b originates from the chiral anomaly induced enhancement of the longitudinal magnetoconductivity, shown in Fig. \ref{Maximum_amplitude_Rotation_angle_for_a_linear_incident_beam}b. The total transmission coefficient depends on external magnetic fields quite strongly and it shows a non-monotonic behavior as a function of frequency at a given magnetic field. On the other hand, the Faraday angle shows a monotonic behavior with an increasing function of the applied magnetic field and decreasing function of the frequency, as shown in Figs. \ref{Maximum_amplitude_Rotation_angle_for_a_linear_incident_beam}c and \ref{Maximum_amplitude_Rotation_angle_for_a_linear_incident_beam}f whereas the Kerr angle shows non-monotonic behaviors in both cases, rather unexpected.

\section{Conclusion}

Light scattering experiments in $\omega < \sigma/\epsilon$ were investigated theoretically in order to prove the axion electrodynamics theory in Weyl metals. We uncovered the existence of longitudinal components in the transmitted/reflected light, which results from longitudinal charge-density fluctuations allowed by the axion electrodynamics. The longitudinal beats in the propagating electric fields are the manifestation of the longitudinal charge density fluctuations. In addition, we found strong dependencies of external magnetic fields in both transmission/reflection coefficients and amount of Faraday/Kerr rotation angles for general configurations. The helicity and amount of Faraday/Kerr rotation angles are determined by $\bm{\nabla} \theta \times \bm{E}_{light} = \bm{B}_{ext} \times \bm{E}_{light}$. Especially, we find various forms of eigenvectors depending on the range of the parameter $\eta$ and a functional dependence of reflectivity under the $\v{E}//\v{B}$ condition, which can be special fingerprints of Weyl metals governed by the axion electrodynamics. Consequently, the light propagation in a Weyl metal phase can be controlled by engineering external magnetic fields.

\section*{ACKNOWLEDGEMENT}

This study was supported by the Ministry of Education, Science, and Technology (No. NRF-2015R1C1A1A01051629 and No. 2011-0030046) of the National Research Foundation of Korea (NRF). KS appreciates helpful discussions with J.-H. Kim.

\end{document}